\documentclass[prl,reprint, twocolumn,showpacs,preprintnumbers,amsmath,amssymb,nofootinbib]{revtex4-1} %,tightenlines
\usepackage{graphicx}

\usepackage{mathrsfs}
\usepackage{hyperref}

\usepackage{color}

\begin{document}

%\title{The four-loop non-planar cusp anomalous dimension in $\mathcal{N} = 4$ SYM} 
 \title{Four-Loop Nonplanar Cusp Anomalous Dimension in $\mathcal{N} = 4$ Supersymmetric Yang-Mills Theory}
 
\author{Rutger H.  Boels}
\email{Rutger.Boels@desy.de}
\affiliation{II. Institut f\"ur Theoretische Physik, Universit\"at Hamburg,  \\ Luruper Chaussee 149, D-22761 Hamburg, Germany}
\author{Tobias Huber}
\email{huber@physik.uni-siegen.de}
\affiliation{Naturwissenschaftlich-Technische Fakult\"at, Universit\"at Siegen, \\ Walter-Flex-Str.~3, 57068 Siegen, Germany}
\author{Gang Yang}
\email{yangg@itp.ac.cn}
\affiliation{CAS Key Laboratory of Theoretical Physics, Institute of Theoretical Physics, \\ Chinese Academy of Sciences, Beijing 100190, China}

\date{\today}

\begin{abstract}
\noindent
The light-like cusp anomalous dimension is a universal function that controls infrared divergences in quite general gauge theories. In the maximally supersymmetric Yang-Mills theory this function is fixed fully by integrability to the three-loop order. At four loops a non-planar correction appears which we obtain for the first time from a numerical computation of the Sudakov form factor. Key ingredients are widely applicable methods to control the number-theoretic aspects of the appearing integrals. Our result shows explicitly that quadratic Casimir scaling breaks down at four loops.%590 characters, including spaces.

\end{abstract}

% \preprint{?}
% 11.25.-w 	Strings and branes
% 11.25.Db	Properties of perturbation theory (strings and branes category)
% 12.38.Bx 	Perturbative calculations (QCD category)



% note 600 character limit for abstract! (includes spaces!)

\maketitle

\section{introduction}
Quantum field theory is used throughout physics due to its unrivalled power to perform systematic computations. Connecting the calculated quantities to observed experimental variables, however, can be complicated. Perturbation theory for the theory of the strong nuclear force, quantum chromodynamics (QCD), for instance can famously only be trusted on very small length scales~\cite{Gross:1973id,Politzer:1973fx}. The physical detectors, however, involve vastly larger length scales where the perturbation theory is guaranteed to break down. A central role in the analysis is played by universal functions, a prime example of which is known as the light-like cusp anomalous dimension (CAD).   

The universality of the CAD makes it appear in many contexts, e.g.~(1) By definition, it refers to the anomalous dimension of a Wilson loop with a light-like cusp~\cite{Polyakov:1980ca, Korchemsky:1985xj}.
(2) It provides the leading infrared (IR) behavior of on-shell amplitudes, see e.g.~\cite{Bern:2005iz}.
(3) It determines the logarithmic growth of anomalous dimensions of high-spin Wilson operators that naturally appear in the operator product expansion description of deep inelastic scattering processes in QCD , see e.g.~\cite{Korchemsky:1988si}.
(4) It is related to the gluon Regge trajectory, and can be obtained from the so-called BFKL equation, see e.g.~\cite{Kotikov:2000pm}.
(5) Via the AdS/CFT correspondence~\cite{Maldacena:1997re}, cusped Wilson operators in the planar limit admit a dual description in terms of spinning strings in a curved background~\cite{Gubser:2002tv}.
There is therefore ample direct physical motivation to compute this function. 

In ${\cal N}=4$ super Yang-Mills (SYM) theory, the planar cusp anomalous dimension is known non-perturbatively via the so-called BES equation~\cite{Beisert:2006ez} based on ideas of the AdS/CFT correspondence and integrability~\cite{Beisert:2010jr}. Consistency with field-theory computations through to four loops has been obtained at the planar level~\cite{Bern:2006ew, Cachazo:2006az, Henn:2013wfa}, and also confirmed by the strong coupling two-loop string computations \cite{Roiban:2007dq}. There is an intriguing connection between ${\cal N}=4$ SYM and QCD results, via the maximal transcendentality property: the ${\cal N}=4$ anomalous dimensions correspond to the ``leading-transcendentality'' contribution in QCD~\cite{Kotikov:2002ab,Kotikov:2004er}. Inspired by this, the three-loop ${\cal N}=4$ results were first obtained in~\cite{Kotikov:2004er} using the three-loop QCD results~\cite{Moch:2004pa}.

In gauge theories with matter fields in the adjoint representation the first correction beyond the planar limit enters into the CAD at four loops. This Letter contains to our knowledge the first computation of this quantity. In perturbation theory four-loop non-planar form factor integrals have to date proven to be largely prohibitively complicated due to the presence of IR divergences. Non-planar corrections are also hard to analyse directly within the AdS/CFT correspondence as they correspond to string loop corrections, and the role of integrability beyond the planar limit is unclear.  Moreover, a quite general conjecture has been posed by extrapolating three-loop results that the non-planar part of the CAD vanishes \cite{Becher:2009qa} in any theory. This is usually called quadratic Casimir scaling of the CAD, and plays an important role in the IR factorization in gauge theories \cite{Gardi:2009qi, Dixon:2009gx, Becher:2009qa, Becher:2009kw, Dixon:2009ur, Ahrens:2012qz}. This scaling is known to break down in planar $\mathcal{N}=4$ SYM at strong coupling \cite{Armoni:2006ux}, as well as through instanton effects \cite{Korchemsky:2017ttd}; see also \cite{Alday:2007mf} for a prediction of violation at four loops. In this Letter the approach will be strictly perturbative.

The minimal scattering-like observable that contains the cusp anomalous dimension is the Sudakov form factor.  In maximally supersymmetric Yang-Mills theory one can use a correlator of a member of the stress-tensor multiplet with two on-shell massless states.  The first computation of the two-loop correction to the Sudakov form factor in $\mathcal{N}=4$ SYM appeared in \cite{vanNeerven:1985ja}.  The three-loop correction to the QCD result was studied in a series of papers~\cite{Baikov:2009bg,Lee:2010cga,Gehrmann:2010ue,Gehrmann:2010tu,vonManteuffel:2015gxa}. In \cite{Gehrmann:2011xn} these results were fine-tuned for the form factor in $\mathcal{N}=4$ SYM to the three-loop order. The integrand for the four-loop Sudakov form factor in $\mathcal{N}=4$ SYM was derived in~\cite{Boels:2012ew} based on the duality between color and kinematics, and its reduction to master integrals was presented in~\cite{Boels:2015yna}. Various other calculations of four-loop corrections in QCD were recently reported~\cite{Henn:2016men,Lee:2016ixa,Ahmed:2017gyt,vonManteuffel:2016xki, Davies:2016jie, Lee:2017mip}. For the five-loop integrand in $\mathcal{N}=4$ SYM  see \cite{Yang:2016ear}.

\section{Review}

\subsection*{Form factor and cusp anomalous dimension}
The Sudakov form factor involves only a single scale $q^2$ which is the Lorentzian norm of the sum of the two massless momenta, i.e. $q^2= (p_1 + p_2)^2$ with $p_1^2=p_2^2=0$. Dimensional analysis and maximal supersymmetry fix the form factor ${\cal F}^{(l)}$  at $l$ loops to be given by
\begin{equation}
{\cal F}^{(l)} = {\cal F}^{\textrm{tree}} g^{2 {l}} (-q^2  )^{-l \epsilon } F^{(l)} \,,
\end{equation}
where the coupling constant is normalised as $g^2 = \frac{g_{\rm YM}^2 C_A}{(4\pi)^2}(4\pi e^{- \gamma_{\text{E}}})^\epsilon$. For a classical Lie-group with Lie-algebra $[ T^{a} ,T^{b} ] = i f^{abc} \, T^c$ and structure constants $f^{abc}$, gauge invariance dictates the color structure to be given by Casimir invariants. Up to three-loop order, only powers $(C_A)^l$ of the quadratic Casimir appear, for which $f^{acd} f^{bcd} = C_{A} \delta^{ab}$ holds. At four loops the quartic invariant $d_{44} = d_A^{abcd}d_A^{abcd}/N_A$ appears in addition to $(C_A)^4$, with $N_A$ the number of generators of the group and 
\begin{align}
d_A^{abcd} =& \frac{1}{ 6} [ f^{\alpha a}{}_{ \beta} f^{\beta b}{}_{ \gamma} f^{\gamma c}{}_{ \delta} f^{\delta d}{}_{ \alpha} + {\text{perms.}(b,c,d)} ] \, . 
\end{align}
Starting from six loops, even higher group invariants appear, see e.g.~\cite{Boels:2012ew}. In $SU(N_c)$, $N_A = N_c^2 -1$, $C_A = N_c$ and $d_{44} = N_c^2/24 \, (N_c^2+36)$ hold.  Without loss of generality, we will focus on $SU(N_c)$ group below.

The form factor has no ultraviolet (UV) divergences since the operator is protected, leaving only IR divergences. If dimensional regularization with $D=4-2 \epsilon$ is used to regulate the latter, $F^{(l)}$ is a purely numerical function of gauge group invariants and $\epsilon$. This function is related to the cusp anomalous dimension $\gamma_{\textrm{cusp}}^{(l)}$ at $l$ loops by~\cite{Bern:2005iz,Mueller:1979ih,Collins:1980ih,Sen:1981sd,Magnea:1990zb},
\begin{align}
(\log F)^{(l)} = -  \bigg[ \frac{\gamma_{\textrm{cusp}}^{({l})} }{(2 {l} \epsilon)^2} + \frac{{\cal G}_{\textrm{coll}}^{({l})} }{2 {l} \epsilon} + {\rm Fin}^{(l)} \bigg] + {\mathcal O}\left(\epsilon\right) \, .
\end{align}
At $l$ loops the planar part $\propto N_c^l$ of $F^{(l)}$ has leading divergence $\propto 1/\epsilon^{2l}$. This function needs to be expanded down to $\epsilon^{-2}$ to extract the $l$-loop CAD, and also higher terms in the Laurent expansion in $\epsilon$ from lower-loop contributions are required. As mentioned, the first occurrence of non-planar (i.e.\ subleading-color) corrections to the CAD is at four loops, due to the appearance of the quartic Casimir invariant $d_{44}$. This invariant therefore breaks quadratic Casimir scaling explicitly. The relation between form factor and cusp anomalous dimension for the non-planar part at four loops is 
\begin{equation}\label{eq:centralrelation}
\left[F^{(4)}\right]_{\textrm{NP}}  = -\frac{\gamma_{\textrm{cusp, NP}}^{(4)}}{(8\epsilon)^2}   + {\mathcal O}\left(\epsilon^{-1}\right) \, ,
\end{equation}
i.e.\ $[F^{(4)}]_{\textrm{NP}}$ has only a double pole in $\epsilon$. Individual integrals that contribute to $[F^{(4)}]_{\textrm{NP}}$ will, however, typically show the full $1/\epsilon^8$ divergence.
The general CAD is believed to be expressible as a rational-coefficient polynomial of Riemann Zeta values $\zeta_n$, and their multi-index generalizations, such as multiple zeta values (MZVs) and Euler sums, see e.g.~\cite{Blumlein:2009cf}. MZVs are denoted by $\zeta_{n_1,n_2,...}$ and have a transcendentality degree which is the sum of their indices, $\sum_{i} n_i$.  At $l$ loops, the planar CAD in ${\cal N}=4$ SYM has uniform transcendentality degree $2l-2$. At four loops for instance, the planar CAD in ${\cal N}=4$ SYM has been computed \cite{Bern:2006ew, Cachazo:2006az, Henn:2013wfa} to be 
 \begin{equation}
 \label{eq:planar-4loop-CAD}
\gamma_{\rm cusp, P}^{(4)}  = - 1752 \zeta_6 - 64\zeta_3^2  \,. 
\end{equation}
We will provide strong evidence that also the non-planar CAD is of uniform transcendentality six at four loops.

In QCD, the known CAD has the same maximal transcendentality degree as in $\mathcal{N}=4$, but also contains lower transcendentality degree constants. The  maximal transcendentality coefficients match between planar $\mathcal{N}=4$ and QCD, an observation known as the maximal transcendentality principle~\cite{Kotikov:2002ab,Kotikov:2004er}.

\subsection*{Integrands, integrals, integral relations}
\begin{figure*}[t]
\includegraphics[width=0.99\textwidth]{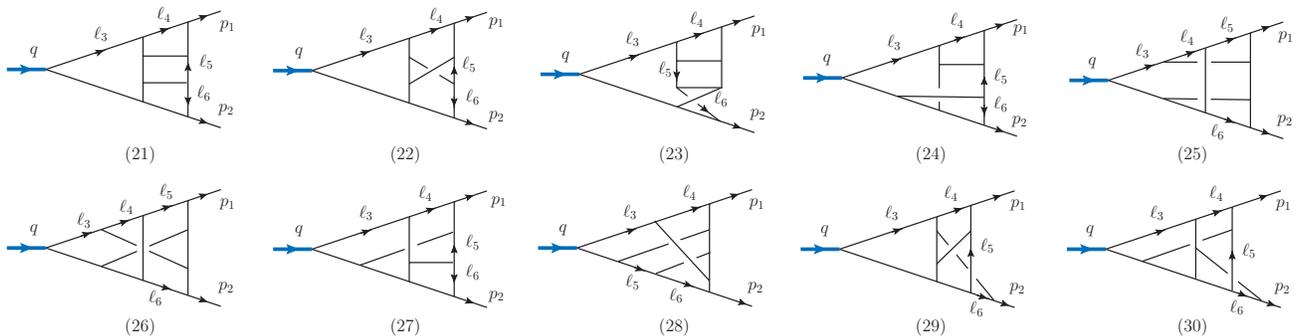}
\caption{\label{fig:NPtops}Integral topologies that contribute to the non-planar form factor at four loops in a basis of UT integrals.}
\end{figure*}

The non-planar part of the Sudakov form factor in $\mathcal{N}=4$ SYM was obtained as a linear combination of a number of four-loop integrals in~\cite{Boels:2012ew} using color-kinematics duality \cite{Bern:2008qj,Bern:2010ue}. The integrals take the generic form
\begin{equation}\label{eq:deffeynmanint}
I = (q^2)^2 \int d^D l_1 \ldots d^D l_4 \; \frac{{N}(l_i, p_j)}{ \prod_{k=1}^{12} D_k } \; ,
\end{equation}
where $D_i$ are propagators and the numerators ${N}(l_i, p_j)$ are quadratic polynomials of Lorentz products of the four independent loop and two independent external on-shell momenta. The explicit expressions of these integrals can be found in~\cite{Boels:2012ew}. There are $14$ distinct integral topologies that contribute to the non-planar CAD, labelled $(21)$~--~$(34)$ in~\cite{Boels:2012ew}, each with 12 internal lines. We will see below that only $10$ of them, $(21)$~--~$(30)$ as shown in Fig.~\ref{fig:NPtops}, contribute to the non-planar form factor if a basis of uniformly transcendental integrals is used.

%%%%%%%%%%%%%%%%%%%%%%%%%%%%%%%%%%%%%%%%%%%%%%%%%%%%%%%%%%%%%%%
%%%%%%%%%%%%%%%%%%%%%%%%%%%%%%%%%%%%%%%%%%%%%%%%%%%%%%%%%%%%%%%

Integrands are only identified up to terms that integrate to zero. Infinitesimal linear reparametrizations of the loop momenta generate such terms, which are known as integration-by-parts (IBP) identities \cite{Chetyrkin:1981qh, Tkachov:1981wb}. With these identities the form factor was simplified in \cite{Boels:2015yna} using the Reduze code~\cite{vonManteuffel:2012np}.  A particular subset of these relations, dubbed `rational IBP' relations and obtained in \cite{Boels:2016bdu}, will play an important role for the problem at hand. Note that integral relations due to graph symmetries are a particular subset of the rational IBP relations. Although simpler integrals emerged in \cite{Boels:2015yna} compared to~\cite{Boels:2012ew}, these have largely evaded integration so far due to their overwhelming complexity.  The obstacle to computing the CAD is therefore to find a complete set of integrals which are simple enough to integrate.

\section{Uniformly transcendental integrals}

The key idea is to find a representation of the four-loop form factor such that all integrals have uniform transcendentality if the dimensional regularization parameter $\epsilon$ is assigned transcendentality $-1$. Such integrals will be referred to as UT integrals. Such a representation of the form factor makes manifest the expected maximal transcendentality property of ${\cal N}=4$ SYM, and has been achieved at three loops in \cite{Gehrmann:2011xn}. Crucially for our purposes, the UT integrals turn out to be much simpler to integrate than generic integrals.

Beyond explicit computation, there are various ways to prove a general integral is a UT integral:  
\begin{itemize}
\item A UT integral can be expressed in the so-called dLog form \cite{Arkani-Hamed:2014via, Bern:2014kca}.
\item A set of integral basis that are all UT integrals can lead to certain simple differential equations \cite{Henn:2013pwa}.
\item The leading singularities, or equivalently, the residues at all poles of a UT integral should be a constant \cite{Bern:2014kca, Bern:2015ple, Henn:2016men}, which is expected to be a sufficient and necessary condition.
\end{itemize}

We have obtained a dLog form, for instance, in topologies $(21)$ and $(23)$ for the integrals with numerator
\begin{equation}
\left[(l_3 - p_1)^2\right]^2 \,
\end{equation} 
each (see Fig.~\ref{fig:NPtops} for the labelling of momenta) and in topology~$(28)$ with numerator
\begin{equation}
(l_3 - l_4- p_2)^2 \times (l_3-p_1)^2 \, .
\end{equation}
The dLog form of the topology-$(21)$ integral as well as its analytic result were also obtained in \cite{Henn:2016men}.

Finding a dLog form is in general a difficult problem. Moreover, the differential equations method is not directly applicable for single-scale problems such as the Sudakov form factor (see \cite{Henn:2013nsa, Henn:2016men} for a work-around). The last criterion in the above bullet-point list, however, allows an algorithm. For this the four-dimensional loop momenta are parametrised as
\begin{equation}
l_1 = \alpha_1 p_1 + \alpha_2 p_2 + \alpha_3 q_1 + \alpha_4 q_2 \, ,
\end{equation}
where $p_1$ and $p_2$ are the external on-shell momenta, and $q_1$, $q_2$ can be chosen as the two complex solutions to
\begin{equation}
q_i^2 =  q_i \cdot p_j = 0 \quad \forall i,j \quad \textrm{and} \quad q_1\cdot q_2 = - p_1\cdot p_2 \,.
\end{equation}
With this, the integration volume changes to
\begin{equation}
\int d^4 l_1 \left( \ldots\right)=  (p_1\cdot p_2)^2  \int \prod_{i=1}^4 d \alpha_i \left( \ldots\right) \, .
\end{equation}
In this so-called parametric form, the leading singularity is studied by computing subsequent residues for all occurring scalar parameters ($16$ at four loops). For this, one needs to choose an order of taking residues in the scalar parameters. If, after taking a residue in a particular parameter, one encounters other than a simple pole in a remaining parameter, the integral is not UT. In principle, there are $16! \sim 2\times 10^{13}$ different orders in which the residues can be taken. We will explain below how we deal with this in practice.

Besides checking UT properties of single integrals, this procedure can be used to constrain the space of potential UT integrals of a given topology.  Start with an Ansatz of a linear combination of momentum products of mass dimension four. Given a set of integrals for a given topology one can see if one or more members fail the UT test for a particular randomly chosen order of residues. If a quadratic or higher pole is found, one can derive a linear constraint on the set of coefficients in the Ansatz to evade the pole. Solving these constraints gives a smaller set of integrals. In the case at hand about a thousand such random checks typically yields a list of candidate UT integrals that do not easily yield further constraints. Through this strategy a set of candidate integrals was obtained for each topology. We note that a related strategy was used in \cite{Bern:2014kca, Bern:2015ple, Henn:2016men}.

For ease of integration it is advantageous to have a form of the numerator as a product of two factors, both quadratic in momenta. To find linear combinations of UT candidate integrals with this property one writes a product-type Ansatz and derives a set of quadratic-constraint equations. In some cases a solution may not exist or may not be useful. For the non-planar four-loop form factor, this happens in topologies $(26)$ and $(25)$, respectively. Here the Ansatz is widened by adding integrals with two fewer propagators, for instance in topology $(25)$ with numerator 
\begin{align}
\left[(p_1 - l_5)^2+2 (l_4-l_5)^2+(l_3-l_4)^2-(l_3-l_5)^2\right. \nonumber \\
\left.-(p_1-l_4)^2 \right]^2-4\, (l_4-l_5)^2 \, (p_1-l_3+l_4-l_5)^2 \, .
\end{align}

Beyond integrals for which a dLog form was obtained, all UT candidates have passed a significant number of residue checks, e.g.\ the two most complicated ones are in topology $(26)$ and have passed $10^5$ such checks. Moreover, as discussed in the next section, the precision of the explicit numerical computation permits a conversion of all results up to ${\cal O}(\epsilon^{-4})$ to analytic expressions which exhibit uniform transcendentality. Taken together, these features provide strong evidence that all candidates are indeed UT integrals.

Having obtained a set of UT candidate integrals, one needs to express the form factor as a linear combination of them. Since the form factor is expected to be uniformly transcendental itself, this linear combination must involve only rational numbers -- not $\epsilon$. The aforementioned `rational IBP' relations obtained in \cite{Boels:2016bdu} prove extremely useful for this task. 

The first result of this Letter is that the non-planar part of the four-loop form factor obtained in~\cite{Boels:2015yna} can be expressed in terms of $23$ UT candidate integrals, providing strong evidence for maximal transcendentality in the non-planar sector. The full list of these integrals and the appropriate linear combination is given in the appendix below, see also~\cite{upcoming}. Interestingly, for topologies $(31)$~--~$(34)$, no UT candidate integrals could be found.

%%%%%%%%%%%%%%%  TABLE  %%%%%%%%%%%%%%%%%%%%%
\begin{table*}[ht]
\caption{Non-planar form factor result and errors. The $\epsilon^{-8}$ entry is afflicted with a rounding error.}
\label{tab:error}
\centering
\begin{tabular}{l | c| c| c| c | c | c | c   } 
\hline\hline
$\epsilon$ order  		& \, $-$8 \, & \, $-$7 \,& \, $-$6 \,& \, $-$5 \,  & \, $-$4 \,  & \, $-$3 \, & \, $-$2 \,    \cr \hline 
result   	& $-3.8\times 10^{-8}$ & $+4.4\times 10^{-9}$ & $-1.2\times 10^{-6}$ & $-1.2\times 10^{-5}$ &  $+3.5\times 10^{-6}$ & $+$~0.0007 & $+$1.56 \cr \hline 
uncertainty    	& $\pm 7.9\times 10^{-9} $ & $\pm 5.7\times 10^{-7}$ & $\pm 1.0\times 10^{-5}$ & $\pm 1.2\times 10^{-4}$ &  $\pm 1.5\times 10^{-3}$  & $\pm$ 0.0186  & $\pm$0.21  \cr \hline \hline
\end{tabular} 
\end{table*}
%%%%%%%%%%%%%%%%%%%%%%%%%%%%%%%%%%%%%%%%%%
\section{Numerical integration}

A variety of techniques exist to compute loop integrals both analytically and numerically, see e.g.~\cite{Smirnov:2012gma} for an overview. Here mainly sector decomposition and Mellin-Barnes (MB) integration have been used, as well as an exact integral result of topology $(21)$ \cite{Henn:2016men}.

Two computer implementations of sector decomposition \cite{Binoth:2000ps} have been used: mostly FIESTA \cite{Smirnov:2008py, Smirnov:2009pb, Smirnov:2013eza, Smirnov:2015mct}, with cross-checks for simpler integrals using SecDec \cite{Carter:2010hi, Borowka:2012yc, Borowka:2015mxa}. An important empirical observation within the sector decomposition approach is the feature that UT integrals usually generate considerably fewer integration terms compared to non-UT siblings of comparable complexity, even though for instance single-sector contributions are not UT separately. The numerical integration with FIESTA is performed using the VEGAS algorithm \cite{Lepage:1980dq} as implemented in the CUBA library \cite{Hahn:2004fe}. When using SecDec, the CUHRE and DIVONNE algorithms are applied.

A second integration technique applied here is MB integration. Several automated tools exist, such as MB tools \cite{Czakon:2005rk, Smirnov:2009up} and AMBRE \cite{Gluza:2007rt, Gluza:2010rn, Blumlein:2014maa}. Given an integral of the form \eqref{eq:deffeynmanint}, the main challenge is to derive valid MB representations for crossed four-loop topologies. Having achieved the latter, the obtained representations are in many cases too high-dimensional to be integrated. Still, efficient MB representations for a subset of (planar and crossed) integrals could be derived by means of a hybrid of the loop-by-loop approach and using the ${\cal F}$ and ${\cal U}$ graph polynomials (see~\cite{upcoming} for details). If an efficient MB representation is available, the numerical precision of the integral is typically 3 to 4 orders of magnitude better compared to sector decomposition.

Before presenting our results, a thorough investigation of the numerical uncertainties is in order. In sector decomposition, several runs with different integration settings were performed. We observe that error bars reported by FIESTA scale with $1/\sqrt{N_s}$, with $N_s$ the number of sampling points (typically in the range of a few up to several $100$ millions). Moreover, we observe that fluctuations upon increasing $N_s$ are well within the reported error bars.
For the leading non-trivial poles, $\epsilon^{- \{8,6,5,4\}}$, the numerical precision in each UT integral is high enough to write the number as a \emph{small} rational multiple of $\{1,\zeta_2,\zeta_3,\zeta_4\}$, respectively. The reliability of the error bars reported by FIESTA was checked by comparing to MB and (semi-)analytic results where available. Altogether, this comparison involves more than seventy data points. The fluctuations for all integrals used are within the reported FIESTA uncertainties, in many cases by over an order of magnitude.

The uncertainties will be conservatively interpreted as the standard deviation of a Gaussian distribution, and errors are added in quadrature.  Our results show no evidence of systematic effects in the FIESTA errors: deviations to other results include both positive and negative signs. A hypothetical systematic error will be modelled by adding uncertainties from individual UT integrals linearly. In conclusion, there is no need to manually inflate the reported uncertainties. Further details will be provided in~\cite{upcoming}.

\subsection{Results}
Our results for the non-planar four-loop form factor up to an overall factor are summarised in table \ref{tab:error}. The first six orders in the $\epsilon$-expansion must vanish by equation\eqref{eq:centralrelation}. The $\epsilon^{-7}$ coefficient should vanish for each integral, and indeed does well within error bars. The coefficients of $\epsilon^{-\{8,6,5,4,3\}}$ can be non-zero in individual integrals but must cancel in the linear combination of the form factor, which is indeed the case, both using direct numerics (see table \ref{tab:error}) as well as the obtained analytic expressions.

The non-trivial coefficient for the non-planar form factor at order $\epsilon^{-2}$ yields the sought-after non-planar four-loop CAD:
\begin{equation}
\label{eq:CAD-result}
\gamma_{\textrm{cusp, NP}}^{(4)}  = -3072\times( 1.56 \pm 0.21 ) \frac{1}{N_c^2} \,,
\end{equation}
where $3072 = 2 \times 24 \times 64 $ is the normalization factor arising from the permutational sum, the color factor~\cite{Boels:2012ew}, and the denominator of~(\ref{eq:centralrelation}), respectively. The significance of a deviation from zero is $~7.4 \sigma$. This is the second major result of this Letter. Adding individual uncertainties linearly would yield $1.56 \pm 0.62$; still significantly non-zero.

Compared to the planar result \eqref{eq:planar-4loop-CAD}: $\gamma_{\rm cusp, P}^{(4)}\sim-1875$, we see that the non-planar CAD has the same sign. If $N_c=3$ is used, its central value becomes $\gamma_{\textrm{cusp, NP}}^{(4)} \sim - 532$, i.e.\ the planar contribution is a factor of 3~--~4 larger.

\section{Discussion}
In this article we present the first computation of the non-planar correction to the cusp anomalous dimension in the maximally supersymmetric Yang-Mills theory, which starts at four loops. 

While integrating a generic set of four-loop form factor master integral remains quite challenging, a key idea leading to the present result is to express the form factor in a basis of uniformly transcendental integrals, which can be constructed algorithmically. While a generic integral contains mixed transcendentality degrees, the UT integrals contain only numbers of fixed (maximal for ${\cal N}=4$ SYM) transcendentality in each order of the $\epsilon$ expansion.  What is interesting and deserves further study is the empirical observation that such simplicity is inherited in sector decomposition, where each sector is no longer UT separately. Once a good set of (candidate) UT basis integrals is determined, the full form factor can be expressed in this basis by using a simple subset of the IBP relations or, in principle, directly by unitarity cuts. In the case at hand, our results provide strong evidence for the maximal transcendentality of the four-loop non-planar form factor in ${\cal N}=4$ SYM. In particular this implies that the four-loop CAD can be written as a rational linear sum of weight-six transcendental numbers.

The UT basis finding algorithm is widely applicable. An immediate interesting application of already obtained results would be to four-loop propagator integrals in QCD, see e.g.~\cite{Ruijl:2017cxj}. A UT basis in QCD would always involve pre-factors with non-negative powers of $\epsilon$, simplifying computations potentially drastically, see \cite{vonManteuffel:2014qoa} for similar ideas. 

Based on the UT basis, a numerical computation has also yielded the first information on the value of the non-planar CAD in $\mathcal{N}=4$, which is statistically significantly non-zero. In particular, our result shows that quadratic Casimir scaling breaks down at four loops in this theory and is therefore not expected to hold in any other theory such as QCD. Direct further research directions include improving precision at order $1/\epsilon^2$ and computing further orders in the $\epsilon$-expansion which contain the so-called collinear anomalous dimension. Computing the four-loop non-planar CAD analytically, also in other theories such as QCD, is a prime further goal.

\begin{acknowledgments}
\section*{Acknowledgements}
The authors would like to thank Dirk Seidel for collaboration in the early stages of this project as well as Bernd Kniehl and Sven-Olaf Moch for discussions and encouragement. This work was supported by the German Science Foundation (DFG) within the Collaborative Research Center 676 ``Particles, Strings and the Early Universe". GY is supported in part by the Chinese Academy of Sciences (CAS) Hundred-Talent Program, by the Key Research Program of Frontier Sciences of CAS, and by Project 11647601 supported by National Natural Science Foundation of China (NSFC).
\end{acknowledgments}

\textbf{Note added:} During the review of this Letter two preprints \cite{Moch:2017uml} and \cite{Grozin:2017css} have appeared which report violation of quadratic Casimir scaling at four loops in QCD.

\appendix
%\begin{widetext}
\section{Supplemental material: \\UT integrals used in the non-planar form factor}

Here we list explicitly the 23 UT integrals $I^{(n)}_{1~-~23}$ that build the non-planar form factor. The superscript $(n)$ indicates the 12 propagators from topology $(n)$ in Fig.~1 of the main article, where also the labelling of the momenta is taken from. In this way, we only have to list the numerator of each integral. Moreover, each integral $I^{(n)}_i$ gets supplemented by a rational pre-factor $c_i$ according to $\vec c = \{1/2, 1/2, 1/2, -1, 1/4, -1/4, -1/4, 2, 1, 4, 1, 1, -1/2, 1, 1, \\ 1, 1, 1, 1, 1, -1, 1/4, 1/2\}$. 
The non-planar form factor is then obtained via $\sum_{i=1,\ldots,23} \, c_i \, I^{(n)}_i$. Integrals $I_{1~-~11}$, $I_{12~-~18}$, and $I_{19~-~23}$, are 12-, 11-, and 10-line integrals, respectively.
\allowdisplaybreaks[1]
\begin{align*}
I^{(21)}_{1}   &=[(\ell_3-p_1)^2]^2\\[0.2em]
I^{(22)}_{2}   &=-(\ell_3-p_1)^2 \, [\ell_4^2+\ell_6^2-\ell_3^2+(\ell_3-\ell_4+p_1)^2\\[0.2em]
               & \qquad \qquad \qquad +(\ell_3-\ell_6-p_1)^2]\\[0.2em]
I^{(23)}_{3}   &=[(\ell_3-p_1)^2]^2\\[0.2em]
I^{(24)}_{4}   &=(\ell_3-p_1)^2 \, [(q-\ell_3-\ell_5)^2 + (\ell_5+p_2)^2 ]\\[0.2em]
I^{(25)}_{5}   &=\left[(p_1 - \ell_5)^2+2 (\ell_4-\ell_5)^2+(\ell_3-\ell_4)^2-(\ell_3-\ell_5)^2 \right. \\[0.2em]
               & \quad \left. -(p_1-\ell_4)^2 \right]^2-4\, (\ell_4-\ell_5)^2 \, (p_1-\ell_3+\ell_4-\ell_5)^2 \\[0.2em]
I^{(26)}_{6}   &=[(\ell_3-\ell_4-\ell_5)^2-(\ell_3-\ell_4-p_1)^2-(\ell_6-p_2)^2-\ell_5^2] \\[0.2em] 
               & \quad \times [\ell_5^2-\ell_4^2-\ell_6^2+(\ell_4-\ell_6)^2] +4\, \ell_5^2 \, (\ell_6-p_2)^2\\[0.2em]
               & \quad + (\ell_4-\ell_5)^2 \, (\ell_3-\ell_4+\ell_6-p_2)^2 \\[0.2em]
I^{(26)}_{7}   &=4\, [(\ell_4-\ell_5) (\ell_3-\ell_4+\ell_5-p_1)]  \\[0.2em]
               & \quad \times [(\ell_4-\ell_6) (\ell_3-\ell_4+\ell_6-p_2)] \\[0.2em]
	       & \quad - (\ell_3-\ell_4)^2 \, (\ell_5+\ell_6-\ell_4)^2 - \ell_5^2 \, (\ell_6-p_2)^2\\[0.2em]
               &  \quad  - 4\, (\ell_4-\ell_5)^2 \, (\ell_3-\ell_4+\ell_6-p_2)^2 - \ell_6^2 \, (\ell_5-p_1)^2 \\[0.2em]
               &  \quad - \ell_4^2 \, (\ell_3-\ell_4+\ell_5+\ell_6-q)^2 \\[0.2em]
I^{(27)}_{8}   &=\frac{1}{2} \left[\ell_3^2 - \ell_4^2 - (\ell_4-\ell_3-p_1)^2\right] \\[0.2em]
               & \qquad \times \left[(\ell_3-\ell_4-\ell_5)^2 + (\ell_5+p_2)^2\right] \\[0.2em]
I^{(28)}_{9}   &=(\ell_3 - \ell_4 - p_2)^2 \, \left[ (\ell_3-\ell_4)^2 - (\ell_3-p_1)^2\right]\\[0.2em]
I^{(29)}_{10}  &=\frac{1}{2} \left[\ell_3^2 - \ell_4^2 - (\ell_4-\ell_3-p_1)^2\right] \, \left[\ell_6 \cdot (\ell_6 - \ell_4 + \ell_3 - p_2)\right]\\[0.2em]
I^{(30)}_{11}  &= (\ell_3-\ell_4-p_2)^2 [(p_1-\ell_4)^2+(\ell_3-\ell_4)^2-(\ell_3-p_1)^2] \\[0.2em]
I^{(27)}_{12}  &= \frac{1}{2} \, (\ell_3-\ell_4)^2 \, \left[2 \, (\ell_4-p_2)^2 + (\ell_6-p_1)^2 - \ell_4^2 + \ell_5^2 \right.  \\[0.2em]
               & \qquad \qquad \qquad \quad\left.- (\ell_4-\ell_6)^2 + 2 \, (p_1+p_2)^2\right] \\[0.2em]
I^{(28)}_{13}  &= \frac{1}{2} \, (\ell_3-\ell_4)^2 \, \left[2 \, (\ell_3-\ell_4-p_2)^2 + (\ell_6-p_1)^2 + \ell_4^2 \right. \\[0.2em]
               & \qquad \qquad \qquad \quad\left.   - (\ell_4-\ell_6)^2 \right] \\[0.2em]
I^{(29)}_{14}  &= (\ell_4-p_1)^2 \, \left[(\ell_3-\ell_4+\ell_6)^2 + (\ell_6-p_2)^2 - \ell_6^2\right] \\[0.2em]
I^{(29)}_{15}  &= \frac{1}{2} \, (\ell_3-p_1-p_2)^2 \, \left[(\ell_4-\ell_6)^2 - (\ell_4-p_2)^2  \right.\\[0.2em]
               & \qquad \qquad \qquad \qquad \quad \left. - (\ell_6-p_1)^2 - (p_1+p_2)^2 \right]\\[0.2em]
I^{(30)}_{16}  &= (\ell_3-p_1-p_2)^2 \, (\ell_5+p_2)^2 \\[0.2em]
I^{(30)}_{17}  &= \frac{1}{2} \, (\ell_4-p_1)^2 \, \left[2\, (\ell_5+p_2)^2 - (\ell_5+p_2+\ell_4-\ell_3)^2 \right] \\[0.2em]
I^{(30)}_{18}  &= \frac{1}{2} \, (\ell_3-\ell_4)^2 \, \left[2\, (\ell_6-\ell_4+p_1)^2 - 3 \, \ell_6^2 \right] \\[0.2em]
I^{(22)}_{19}  &= (\ell_3-\ell_4)^2 \,  (p_1-\ell_3+\ell_6)^2 \\[0.2em]
I^{(22)}_{20}  &= \ell_6^2 \,  (p_1-\ell_4)^2 \\[0.2em]
I^{(24)}_{21}  &= (p_1 - \ell_3-\ell_5)^2 \,  (\ell_3-p_1-p_2)^2 \\[0.2em]
I^{(24)}_{22}  &= \ell_5^2 \,  (\ell_3-p_1-p_2)^2 \\[0.2em]
I^{(28)}_{23}  &= (\ell_4 - p_1)^2 \,  (\ell_3-\ell_4+\ell_5-p_2)^2 \, .
\end{align*}

\bibliographystyle{apsrev4-1}

%\bibliography{References}


\end{document}